%% file: main.tex
\documentclass[sigconf,natbib,screen=true]{acmart}

\input{packages}
\input{copyright}

\input{definitions}
\input{authors}

\acmSubmissionID{1151}

\title[Propensity Estimation for Counterfactual Learning to Rank]{Cascade Model-based Propensity Estimation\\ for Counterfactual Learning to Rank}

\allowdisplaybreaks

\begin{document}

\begin{abstract}
	Unbiased \ac{CLTR} requires click propensities to compensate for the difference between user clicks and true relevance of search results {via \ac{IPS}}.
	Current propensity estimation methods assume that user click behavior follows the \ac{PBM} and estimate click propensities based on this assumption. 
	However, in reality, user clicks often follow the \ac{CM}, where users scan search results from top to bottom and where each next click depends on the previous one.
	In this cascade scenario, \ac{PBM}-based estimates of propensities are not accurate, which, in turn, hurts \ac{CLTR} performance.
	In this paper, we propose a propensity estimation method for the cascade scenario, called \OurMethod.
	We show that \OurMethod{} keeps \ac{CLTR} performance close to the full-information performance in case the user clicks follow the \ac{CM},
	while \ac{PBM}-based \ac{CLTR} has a significant gap towards the full-information.
	The opposite is true if the user clicks follow \ac{PBM} instead of the \ac{CM}.
	Finally, we suggest a way to select between \ac{CM}- and \ac{PBM}-based propensity estimation methods based on historical user clicks.
\end{abstract}

\maketitle

\acresetall

\input{sections/01-Introduction}
\input{sections/03-RelatedWork}

\input{sections/04-Method}
\input{sections/05-ExperimentalSetup}
\input{sections/06-Implementation}
\input{sections/07-Results}
\input{sections/08-Conclusion}

\section*{Code and data}
To facilitate the reproducibility of the reported results this work only made use of publicly available data and our experimental implementation is publicly available at \url{https://github.com/AliVard/CM-IPS-SIGIR20}.

\begin{acks}
This research was partially supported by
the Netherlands Organisation for Scientific Research (NWO)
under pro\-ject nr
652.\-002.\-001, 
and
the Innovation Center for Artificial Intelligence (ICAI).
All content represents the opinion of the authors, which is not necessarily shared or endorsed by their respective employers and/or sponsors.
\end{acks}

\bibliographystyle{ACM-Reference-Format}
\bibliography{references}

\end{document}

%% file: packages.tex
\PassOptionsToPackage{dvipsnames}{xcolor}

\usepackage{dsfont}
\usepackage{acronym}
\usepackage[noend]{algorithmic}
\usepackage{algorithm}
\usepackage{bbm}
\usepackage{beramono}
\usepackage{bm}
\usepackage{booktabs}
\usepackage[skip=0pt]{caption}
\usepackage{cleveref}
\usepackage[T1]{fontenc}
\usepackage{graphicx}
\usepackage{listings}
\usepackage{multirow}
\usepackage{natbib}
\usepackage{subcaption}
\usepackage{tabularx}
\usepackage[dvipsnames]{xcolor}
\usepackage{enumerate}
\usepackage[inline]{enumitem}
\usepackage{numprint}
\usepackage{hyperref}
\usepackage[most]{tcolorbox}
\usepackage{outlines}
\usepackage{mathrsfs}
\usepackage{mathtools}
\usepackage{amsthm}

\usepackage{adjustbox}
\usepackage{array}

%% file: copyright.tex
\copyrightyear{2020}
\acmYear{2020}
\setcopyright{acmlicensed}\acmConference[SIGIR '20]{Proceedings of the 43rd International ACM SIGIR Conference on Research and Development in Information Retrieval}{July 25--30, 2020}{Virtual Event, China} \acmBooktitle{Proceedings of the 43rd International ACM SIGIR Conference on Research and Development in Information Retrieval (SIGIR '20), July 25--30, 2020, Virtual Event, China}
\acmPrice{15.00}
\acmDOI{10.1145/3397271.3401299} \acmISBN{978-1-4503-8016-4/20/07}

%% file: definitions.tex
\definecolor{av}{RGB}{0, 150, 0}
\definecolor{mdr}{RGB}{200, 0, 0}

\newcommand{\OurMethod}{\ac{CM-IPS}}

\acrodef{IR}{information retrieval}
\acrodef{LTR}{learning to rank}
\acrodef{ARP}{average relevance position}
\acrodef{DCG}{discounted cumulative gain}
\acrodef{EM}{expectation maximization}
\acrodef{IPS}{inverse propensity scoring}
\acrodef{CLTR}{counterfactual learning to rank}
\acrodef{PBM}{position-based click model}
\acrodef{CM}{cascade model}
\acrodef{IPS}{inverse propensity scoring}
\acrodef{CM-IPS}{cascade model-based inverse propensity scoring}
\acrodef{PBM-IPS}{position-based model-based inverse propensity scoring}
\acrodef{SERP}{search engine result page}
\acrodef{DCM}{dependent click model}
\acrodef{DBN}{dynamic bayesian network}
\acrodef{CCM}{click chain model}
\acrodef{MLE}{maximum likelihood estimation}
\acrodef{CBM}{cascade-based models}
\theoremstyle{definition}

\theoremstyle{definition}

\allowdisplaybreaks

\theoremstyle{remark}

\tcbset{
	frame code={}
	left=0pt,
	right=0pt,
	top=0pt,
	bottom=0pt,
	colback=gray!50,
	colframe=white,
	width=\dimexpr\textwidth\relax,
	enlarge left by=0mm,
	boxsep=5pt,
	arc=4pt,outer arc=0pt,
}

\hypersetup{
	colorlinks,
	citecolor=black,
	filecolor=black,
	linkcolor=black,
	urlcolor=black
}

\newcolumntype{R}[2]{%
    >{\adjustbox{angle=#1,lap=\width-(#2)}\bgroup}%
    l%
    <{\egroup}%
}
\newcommand*\rot{\multicolumn{1}{R{55}{1.9em}}}

%% file: authors.tex
\author{Ali Vardasbi}
\affiliation{%
  \institution{University of Amsterdam}
  \city{Amsterdam}
  \country{The Netherlands}
}
\email{a.vardasbi@uva.nl}

\author{Maarten de Rijke}
\affiliation{%
  \institution{\mbox{}\hspace*{-4mm}\mbox{University of Amsterdam \& Ahold Delhaize}}
  \city{Amsterdam}
  \country{The Netherlands}
}
\email{m.derijke@uva.nl}

\author{Ilya Markov}
\affiliation{%
  \institution{University of Amsterdam}
  \city{Amsterdam}
  \country{The Netherlands}
}  
\email{i.markov@uva.nl}

%% file: sections/01-Introduction.tex

\section{Introduction}

Traditional \ac{LTR} and online \ac{LTR} require explicit relevance labels and intervention through search engine results, respectively \cite{liu2009learning, oosterhuis2018differentiable}.
In contrast, \ac{CLTR} only requires historical click logs for learning.
Obtaining and using historical click logs incurs no extra cost and does not impose any risk of reduced user satisfaction.
More importantly, such benefits come without any significant reduction in \ac{LTR} performance~\cite{joachims2017unbiased,wang2016personal,ai2018unbiased,wang2018position}.
However, user clicks are known to suffer from different types of bias, such as position bias, selection bias, trust bias, etc.~\cite{chuklin2015click}.
Due to these types of bias, each result on a \ac{SERP} has a different \textit{propensity} of being clicked.
Since \ac{CLTR} learns from user clicks, it should take those propensities into account.
To make \ac{CLTR} unbiased, the \ac{IPS} method has been introduced in~\cite{wang2016personal,joachims2017unbiased}.

In \ac{IPS}-based \ac{CLTR}, click models are used to estimate propensities~\cite{joachims2017unbiased}.
Even though the theoretical \ac{IPS} method does not rely on any specific click model~\cite{joachims2017unbiased}, current \ac{IPS}-based \ac{CLTR} experiments rely on the \ac{PBM}~\cite{joachims2017unbiased,wang2016personal,ai2018unbiased,wang2018position}.
In \ac{PBM}, the probability of examining a result depends on the result's rank only and not on any other context, such as clicks on other items.

Although \ac{PBM} is a well-performing click model~\cite{chuklin2015click}, it does not always approximate user clicks well~\cite{grotov2015comparative}.
Importantly, \ac{PBM} fails to represent the cascade user click behavior, where a user scans a \ac{SERP} from top to bottom and where each next click depends on the previous click~\cite{craswell2008experimental} -- such behavior is often observed in practice~\cite{chuklin2015click,grotov2015comparative}.
In this case, \ac{PBM}-\ac{IPS} estimators are not accurate and \ac{CLTR} performance drops considerably.
Look, for example, at Figure~\ref{fig:pbm_vs_dcm_intro}.
{\renewcommand{\arraystretch}{0.01}
\begin{figure}[b]
	\centering	
\begin{tabular}{l c}
\rotatebox[origin=lt]{90}{\hspace{2.25em} \small nDCG@10} &
	\includegraphics[width=0.4\textwidth]{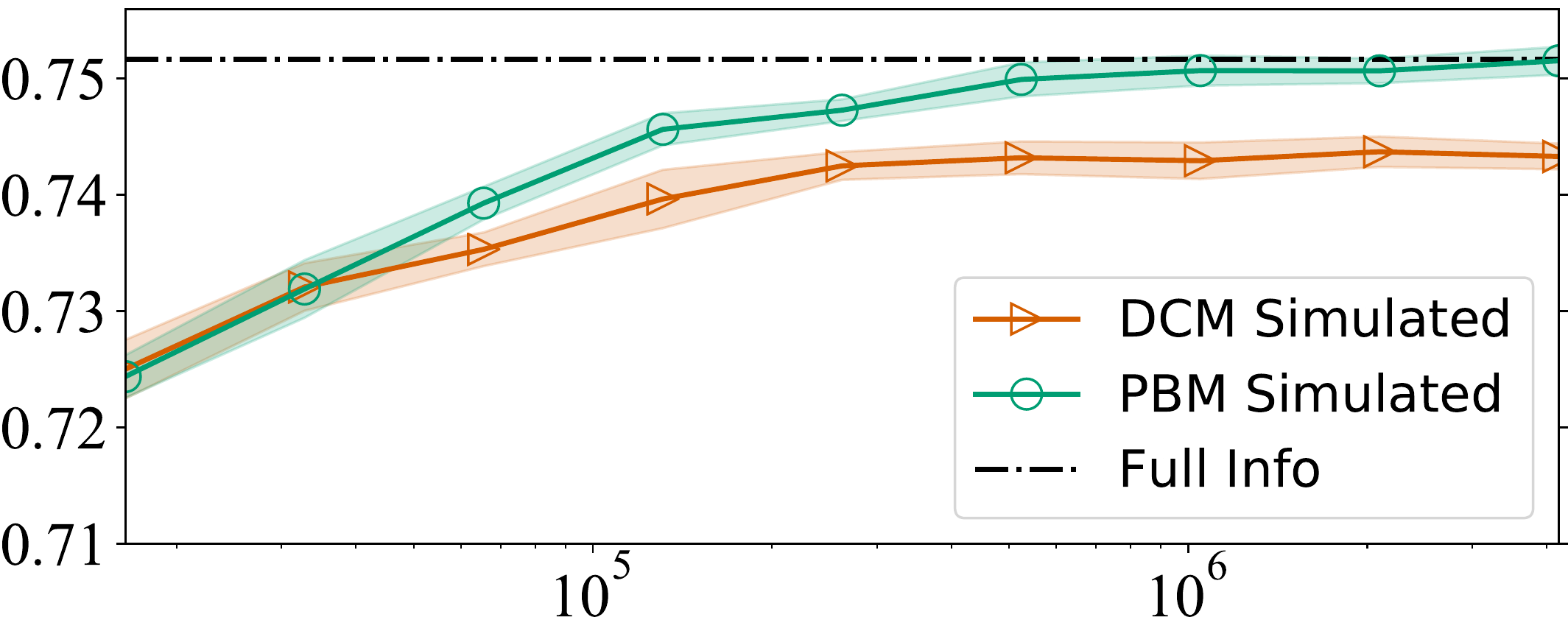} \\
& \small Number of Training Clicks
\end{tabular}
	\caption{Performance of PBM-IPS CLTR on different sets of simulated clicks.}
	\label{fig:pbm_vs_dcm_intro}
\end{figure}
}
There, the \ac{PBM}-\ac{IPS} \ac{CLTR} is trained over two different simulated click logs: one following \ac{PBM} and the other following \ac{DCM}~\cite{DCM2009} one of the popular cascade-based models.
When trained on the same number of clicks and the same queries, \ac{PBM}-\ac{IPS} performs significantly better on the \ac{PBM} than \ac{DCM} simulated clicks. 
The ``Full Info'' legend in this plot shows the performance of LTR trained on real relevance tags instead of simulated clicks.
\ac{PBM}-\ac{IPS} performs close to the full-info only when trained on \ac{PBM} simulated clicks.

In this paper, we first experimentally validate this observation for different parameter settings.
We apply \ac{PBM}-\ac{IPS} unbiased \ac{CLTR} on various sets of simulated clicks and show that \ac{PBM}-\ac{IPS} \ac{CLTR} only performs well when the simulated clicks are drawn based on the \ac{PBM}.
The significance of these results is also in noticing that the current \ac{IPS} unbiased \ac{CLTR} papers all use \ac{PBM}-\ac{IPS} estimation in their experiments~\cite{joachims2017unbiased,ai2018unbiased,wang2018position}.
To fill in the gaps of the \ac{PBM}-based \ac{CLTR} performance on cascade-based clicks, we provide \OurMethod~ and derive closed form formulas for click propensities in three widely used cascade-based click models, \ac{DCM}~\cite{DCM2009}, \ac{DBN}~\cite{DBN2009}, and \ac{CCM}~\cite{CCM2009}.
We experimentally show the effectiveness of our derived propensity formulas for \ac{DCM}.

To sum up, in this paper we are interested in the following research questions:
\begin{enumerate*}[label=(RQ\arabic*)]
	\item Can a \ac{PBM}-\ac{IPS} \ac{CLTR} effectively learn from clicks when the user click behavior is closer to cascade-based models?
	\label{rq:pbm_effectiveness} 
	\item What are the cascade model-based propensity alternatives which are more suitable for \ac{IPS} \ac{CLTR} in the presence of cascade user click behavior?
	\label{rq:cascade_based_propensity} 
\end{enumerate*}
\input{sections/tables/notations.tex}
Table~\ref{tab:notations} summarizes the notation we use in the paper. 

%% file: sections/tables/notations.tex
\begin{table}
    \caption{Notation.}
    \label{tab:notations}
    \begin{tabular}{l  l}
    \toprule
        Parameter~ 
        & Description \\
    \midrule
        $x_j$ & representation of query, document pair at position $j$\\
        $c_j\in\{0,1\}$ & click on result $x_j$ at position $j$ \\
        $r_j\in\{0,1\}$ & relevance of result $x_j$ at position $j$ \\
        $e_j\in\{0,1\}$ & examination of result $x_j$ at position $j$ \\
        $\mathcal{X}_q$ & ordered set of the results corresponding query $q$ \\
        $q_S$ & the query of session $S$ \\
    \bottomrule
    \end{tabular}
\end{table}

%% file: sections/03-RelatedWork.tex

\section{Related Work}

\paragraph{Click models}
Click models model user behavior.
Most click models factorize the click probability into two independent probabilities: the probability of examination and the probability of attractiveness (or relevance)~\cite{chuklin2015click}.
In order to predict the examination probability, various probabilistic click models with different assumptions have been proposed. 
The \acfi{PBM} assumes that the examination probability of a result only depends on its rank in the result list. 
There are several cascade-based models that assume that a user examines the results on the \ac{SERP} linearly, from top to bottom, until she is satisfied with a result and abandons the session.
See Section~\ref{sctn:method} for details.
The true click model of a given (set of) click log(s) is not known, but unbiased \ac{CLTR} requires the knowledge of the click propensities.
Consequently, a click model is usually assumed for the click logs and the click propensities are estimated based on that assumed click model \cite{joachims2017unbiased,wang2018position}.

\paragraph{Learning PBM Propensities}
In LTR, it is common to optimize the sum of some performance metric only over relevant training documents \cite{joachims2017unbiased,ai2018unbiased,agarwal2019addressing}. 
However, in the click logs, the $r_x$ are unknown. 
What is observed is $c_x$. 
According to the examination hypothesis, clicks appear on the relevant results that are also \textit{examined}. 
Hence, the click signals are biased by the examination probability. 
To debias these signals, \citet{joachims2017unbiased} propose to use the so-called \acfi{IPS} method:
\vspace*{-0.5mm}
\begin{equation}
	\label{eqn:unbiased_loss}
	\hat{\mathscr{L}}_{S}=\sum_{x_j\in\mathcal{X}_{q_S}} \frac{ c_j^{(S)} \cdot \mathcal{L}_{x_j} }{P\left(E_j=1 \mid q_S\right)},
\end{equation}
\vspace*{-0.5mm}\noindent%
where $P\left(E_j=1 \mid q_S\right)$ is the marginalized examination probability over all the sessions with the same query:
\vspace*{-0.5mm}
\begin{equation}
	\label{eqn:marginalized_examination}
	P(E_j=1 \mid q)=E_{S \mid q_S=q}\left[P\left(E_j=1 \mid S\right)\right].
\end{equation}
\vspace*{-0.5mm}\noindent%
So far, \ac{LTR} models dealing with click signals assume \ac{PBM} for the clicks (in practice) and estimate the propensities based on this assumption~\cite{ai2018unbiased,joachims2017unbiased}.
In \ac{PBM} 
one can write:
\vspace*{-0.5mm}
\begin{equation}
	\label{eqn:PBM}
	P_{\text{PBM}}(E_{j}=1 \mid q)=P_{\text{PBM}}(E_{j}=1)=\theta_j
\end{equation}

\vspace*{-0.5mm}\noindent%
Existing \ac{CLTR} work builds on the \ac{PBM} assumption~\citep{ai2018unbiased,wang2018position}.
But \ac{PBM} is not necessarily the best fitting model in all situations.
\paragraph{Cascade Bias}
\citet{chandar2018estimating} discuss the idea that the existing \ac{CLTR} methods do not consider the cascade bias, i.e. higher ranked relevancy dependent examination of items.
They focus on counterfactual evaluation of rankers and show that, in presence of cascade bias, a \textit{Context-Aware} \ac{IPS} ranker has a higher Kendall's tau correlation with the full information ranker than that of a simple \ac{IPS}.
Though the basic ideas of~\citep{chandar2018estimating} are the same as the current paper, there at least four important differences:
\begin{enumerate*}[label={(\roman{*})}]
	\item We propose closed form propensity formulas for cascade models, while they directly estimate the propensities using result randomization.
	\item We employ~\OurMethod~ in \ac{CLTR} to learn the ranker, as opposed to evaluating the rankers.
	\item Unlike them, we prove that the hidden click probabilities can be replaced with observed clicks without violating the unbiasedness.
	\item We use real query-document features for training our \ac{CLTR}, whereas they only use fully simulated features in their experiments.
\end{enumerate*}

%% file: sections/04-Method.tex

\section{Cascade Model-based Propensity Estimation}
\label{sctn:method}
We derive recursive formulas for propensity estimation in popular cascade models based on clicks on a query session.
We use \OurMethod~ to refer to an \ac{IPS} method that uses these formulas.
For each of \ac{DCM}, \ac{DBN} and \ac{CCM}, we derive the examination probability at a position, based on the model parameters and the clicks over the previous positions.
This exercise is not necessary for PBM since the propensities are the parameters themselves and examination at a position is independent of user behavior on other positions.

Before proceeding to specific propensity formulas for each click model, we need to rewrite the original IPS method proposed in \cite{joachims2017unbiased} to make it more suitable for \ac{CBM}.
Let us define the IPS per query loss as $\hat{\mathscr{L}}_{q}=\sum_{S \mid q_S=q} \hat{\mathscr{L}}_{S}$.
In what follows we show that, in \ac{CBM}, if the marginalized $P(E_j=1 \mid q)$ in (\ref{eqn:unbiased_loss}) is replaced with the session dependent probabilities $P(E_j=1 \mid C_{<j})$, the per query loss will remain asymptotically unchanged.
For brevity, we will drop the summation over positions as well as the $q_S=q$ condition.
\begin{equation}
	\label{eqn:proof}
\begin{split}
	\hat{ \mathcal{L} }_q[j] = & \sum_{S} \frac{ c_j^{(S)} \cdot \mathcal{L}_{x_j} }{P\left(E_j=1 \mid q\right)} = \frac{ \mathcal{L}_{x_j} }{P\left(E_j=1 \mid q\right)} \sum_{S} c_j^{(S)} \\
	= & N_q \cdot \frac{ \mathcal{L}_{x_j} \cdot P(C_j=1 \mid q) }{P\left(E_j=1 \mid q\right)} = N_q \cdot P(R_j=1) \cdot \mathcal{L}_{x_j} \\
	= & \sum_{S} \frac{ P(C_j=1 \mid c_{<j}^{(S)}) }{ P(E_j=1 \mid c_{<j}^{(S)})} \cdot \mathcal{L}_{x_j} 
	\stackrel{(\ref{eqn:empirical_expectation})}{=}  \sum_{S} \frac{ c_j^{(S)}  \cdot \mathcal{L}_{x_j}}{ P(E_j=1 \mid c_{<j}^{(S)})}
\end{split}
\end{equation}
where the last equality is empirically valid based on Eq.~(\ref{eqn:empirical_expectation}) below.
In \ac{CBM}, we can write for a general function $g$ depending on the clicks before, and including, position $j$:
\begin{align}
	& \sum_{S} P(C_j=1 \mid c_{<j})\cdot g(c_{\leq j}) \nonumber \\
	= & \sum_{c_{<j}\in \{0,1\}^{j-1}} |S_{c_{<j}}| \cdot P(C_j=1 \mid c_{<j}) \cdot g(c_{\leq j}) \nonumber \\
	\simeq & \sum_{c_{<j}\in \{0,1\}^{j-1}} \sum_{S \in S_{c_{<j}}} c_j^{(S)} \cdot g(c_{\leq j}) = \sum_{S} c_j^{(S)} \cdot g(c_{\leq j}) 
		\label{eqn:empirical_expectation}
\end{align}
where $S_{c_{<j}}=\{S \mid c_{<j}^{(S)}=c_{<j}\}$ and the third line is the empirical estimation of the second line.
For \ac{CBM}, the marginalized $P(E_j=1)$ depends on the relevance probabilities of higher ranked results.
But relevance is unknown during the CLTR and is yet to be learned.
Instead of using EM algorithms to estimate relevance and marginalized examination probabilities, we propose to simply use the $P(E_j=1 \mid C_{<j})$ which has been shown here to be empirically equivalent to the original loss.

Next we will derive separate formulas for $P(E_j=1 \mid C_{<j})$ in DCM, DBN and CCM models.
\paragraph{DCM}
In DCM, the user examines the results from top to bottom until she finds an attractive result, $P(E_{j+1}=1\mid E_{j}=1,C_{j}=0)=1$.
After each click, there is a position dependent chance that the user is not satisfied,
$P(E_{j+1}=1\mid C_{j}=1)=\lambda_j$.
Therefore:
\begin{equation}
	\label{eqn:DCM}
	P_{\text{DCM}}(E_{j}=1 \mid c_{<j})=\prod_{i<j}(1-c_{i}(1-\lambda_i))
\end{equation}
\vspace*{-3.5mm}
\paragraph{DBN}
In DBN, 
there is another binary variable to model the user's satisfaction after a click.
A satisfied user abandons the session, $P(E_{i+1}=1 \mid S_i=1)=0$.
An unsatisfied user may also abandon the session with a constant probability $\gamma$.
Finally, after a click, the satisfaction probability depends on the document, $P(S_i=1 \mid C_i=1)=s_{x_i}$.
Thanks to Eq.~(\ref{eqn:proof}), we only need the session specific examination probability, which can be derived as follows:
\begin{equation}
	\label{eqn:DBN}
	P_{\text{DBN}}(E_{j}=1 \mid c_{<j})=\prod_{i<j} \gamma \cdot (1-c_{i} \cdot s_{x_i})
\end{equation}
\paragraph{CCM}
The CCM is a generalization of DCM where continuing to examine the results before a click is not deterministic, $P(E_{j+1}=1 \mid E_j=1, C_j=0)=\alpha_1$.
The probability of continuing after a click is not position dependent, but relevance dependent, $P(E_{j+1} \mid C_j=1)=\alpha_2 (1-R_i) + \alpha_3 R_i$.
Similar to \ac{DCM} we have:
\begin{equation}
	\label{eqn:CCM}
	\begin{split}	
		P_{\text{CCM}}&(E_{j}=1 \mid c_{<j})=  \prod_{i<j}(\alpha_1-c_{i}(\alpha_1 - \alpha_2 (1-R_i) - \alpha_3 R_i)).
	\end{split}	
\end{equation}

\paragraph{Parameter Estimation}
\label{sctn:param_estimate}
In click model studies, parameter estimation is performed for each query over the sessions initiated by that query~\cite{chuklin2015click}:
a low variance estimation requires a great number of sessions for each query.
In CLTR studies, on the other hand, one uses features of query-document pairs in order to generalize well to tail queries~\cite{wang2018position}.
We leave the feature-based parameter estimation of \ac{CBM} as future work.

%% file: sections/05-ExperimentalSetup.tex

\vspace*{-1.5mm}
\section{Experimental Setup}

\paragraph{Dataset}
We use the Yahoo! Webscope~\cite{chapelle2011yahoo} dataset for LTR with synthetic clicks.
Our methodology follows previous unbiased LTR papers~\cite{joachims2017unbiased,ai2018unbiased}.
We use binary relevance, considering the two most relevant levels as $r=1$.
We randomly select 50 queries from the training set and train a LambdaMART model over them to act as the initial ranker.
The documents of all the queries are ranked using this initial ranker and the top 20 documents are shown to the virtual user.
We remove all the queries which have no relevant documents in their top 20 documents.
Consequently, the train and test sets have 11,474 and 4,085 queries, respectively.
User behavior is modeled by \ac{PBM} or \ac{DCM} with various parameter assignments (see below).
Sessions with at least one click are kept in the training set.

The reported results use $4M$ clicks for training, where the performance of \ac{CLTR} is converged.

\paragraph{Click simulation}
\label{sctn:experiments_click_models}
We use PBM and DCM for generating click data.
For PBM, we use the widely used reciprocal formula for the examination probability \cite{joachims2017unbiased,ai2018unbiased} (see Eq.~(\ref{eqn:PBM})):
\begin{equation}
    \label{eqn:PBM_theta}
    P_{\text{PBM}}(E_{j}=1)=\theta_j=\left(\frac{1}{j}\right)^\eta,
\end{equation}
\noindent
with $\eta\in \left\{0.5,1,2\right\}$.

For DCM, we use a similar formula for $\lambda$
(see Eq.~(\ref{eqn:DCM})):
\begin{equation}
    \label{eqn:DCM_lambda}
    P_{\text{DCM}}(E_{j+1}=1 \mid C_{j}=1)=\lambda_j=\beta \left(\frac{1}{j}\right)^\eta,
\end{equation}
\noindent
where $\beta$ and $\eta$ are tuning parameters. 
We use $\beta \in \left\{0.6,1\right\}$ and $\eta\in \left\{0.5,1,2\right\}$. 

In both \ac{PBM} and \ac{DCM} cases, we used a noise (i.e. click on examined non-relevant items) with probability $0.05$.

\paragraph{Experimental protocol}
\label{sctn:pbm_effectiveness_protocol}
To investigate the effectiveness of \ac{PBM}-\ac{IPS} as well as~\OurMethod, we try to train a \ac{CLTR} over different sets of simulated click logs as explained above.
We use DLA~\cite{ai2018unbiased} to learn the click propensities based on the PBM assumption and MLE~\cite{DCM2009} to estimate $\lambda$'s for \ac{DCM}.
Similar to other works on CLTR, we evaluate the rankings using explicit relevance judgements in the test set.
We use nDCG at 10 to compare the rankings.
We also report full-information results where the true relevance labels are used for training,  i.e., the highest possible performance (skyline).

%% file: sections/06-Implementation.tex

\paragraph{LTR implementation}
Different LTR algorithms have been used for CLTR, including SVMRank \cite{joachims2017unbiased}, neural networks (NNs) \cite{ai2018unbiased,ai2019learning}, and LambdaMART \cite{wang2018position}. 
The differences are minimal \cite{ai2018unbiased}.
We follow \cite{ai2018unbiased} and model the score function by a DNN, with the loss being softmax cross entropy.
We use three layers with sizes $\left\{512,256,128\right\}$ and $elu$ activation; 
the last two layers use dropout with a dropping probability of $0.1$. 
Based on~\citep{swaminathan2015batch} we use a propensity clipping constant of $100$ to avoid exploding variance.

%% file: sections/07-Results.tex
 \vspace*{-2mm}
\section{Results}

\paragraph{CM-IPS effectiveness}
\label{sctn:pbm-effectiveness_results}
In order to analyze the effectiveness of \OurMethod, we follow the protocol described in Section~\ref{sctn:pbm_effectiveness_protocol}.
Fig.~\ref{fig:comparison} shows the performance of \OurMethod~compared to \ac{PBM}-\ac{IPS} \ac{CLTR} on numerous simulated click sets.

{\renewcommand{\arraystretch}{0.01}
 \begin{figure}[t]
 	\centering	
 \begin{tabular}{l c c c c c c c c c c c}
 \rotatebox[origin=lt]{90}{\hspace{1.25em} \small nDCG@10} &
 \multicolumn{10}{c}{
 	\includegraphics[width=0.445\textwidth]{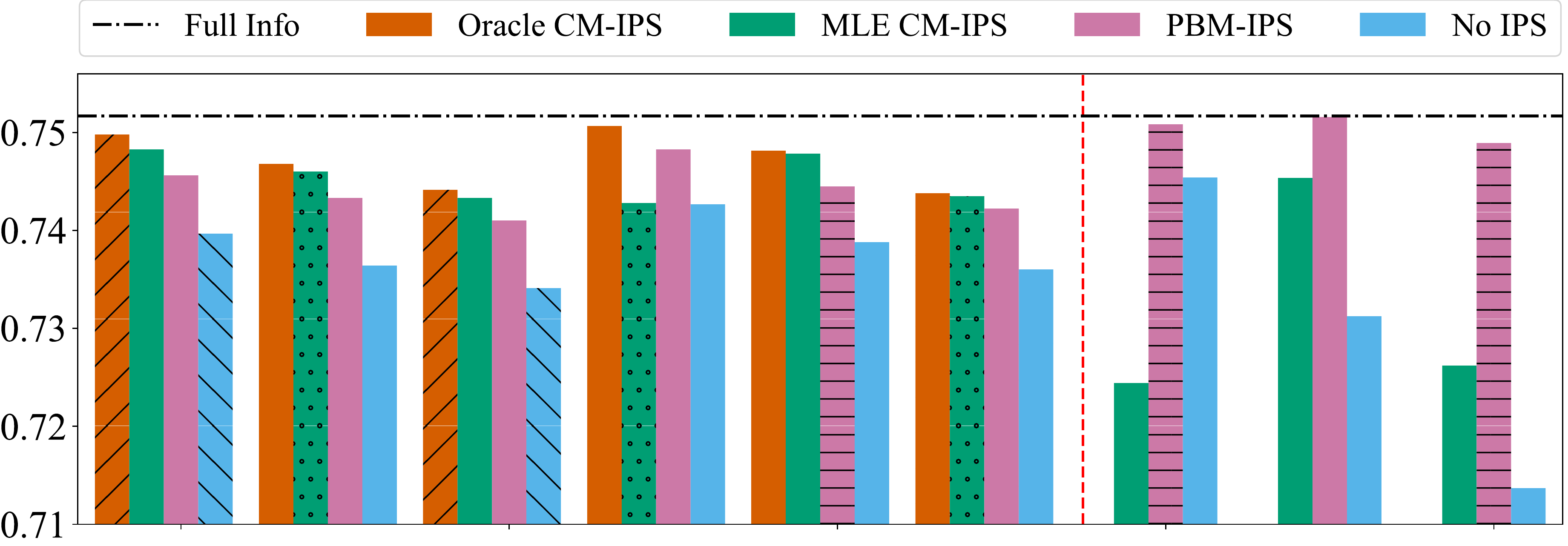} 
 } \\
&
\rot{\small dcm\_0.6\_0.5}&
\rot{\small dcm\_0.6\_1.0}&
\rot{\small dcm\_0.6\_2.0}&
\rot{\small dcm\_1.0\_0.5}&
\rot{\small dcm\_1.0\_1.0}&
\rot{\small dcm\_1.0\_2.0}&
&
\rot{\small \hspace{1em} pbm\_0.5}&
\rot{\small \hspace{1em} pbm\_1.0}&
\rot{\small \hspace{1em} pbm\_2.0}
 \end{tabular}
 	\caption{Performance of PBM-IPS and CM-IPS CLTR on different sets of simulated clicks. We repeated each experiment 15 times and report the mean value.}
 	\label{fig:comparison}
 \end{figure}
 }

The x-axis shows the method used for simulating clicks: either ``$\text{dcm}\_\beta\_\eta$'' or ``$\text{pbm}\_\eta$'' as explained in Section~\ref{sctn:experiments_click_models}.
The y-axis is the ranking performance of \ac{CLTR} methods in terms of nDCG at $10$.
We see both \ac{PBM}-\ac{IPS} and \OurMethod{} improve the biased na\"ive \ac{LTR} (indicated by ``No IPS'') in all cases.
When using \OurMethod{} correction with oracle parameters for \ac{DCM} simulated click sets (on the left of the vertical dashed line), the performance is consistently improved compared to \ac{PBM}-\ac{IPS}.
All the differences are significant with $p<0.0001$ except for $\text{dcm}\_1.0\_2.0$ which has $p<0.05$.
The reverse holds for \ac{PBM}-\ac{IPS}: it has a better performance for \ac{PBM} simulated click sets (on the right of the vertical dashed line), compared to \OurMethod.
In these datasets, the performance of \ac{PBM}-\ac{IPS} \ac{CLTR} is very close to the full-information case.
These observations suggest that for a great variety of different parameter settings of \ac{DCM}, the \ac{PBM}-\ac{IPS} correction cannot remove the bias, while \OurMethod{} can.
More generally, Fig.~\ref{fig:comparison} shows that when the click behavior and the correction method agree, the results are consistently better than the other case.

There is one practical issue that we leave as future work.
This concerns the parameter estimation of \ac{DCM}.
The above discussions are valid when using the oracle parameter values for $\lambda_j$.
It is worth mentioning that, unlike \ac{PBM}, the parameters and the propensities are two different things in \OurMethod.
The novelty of \OurMethod{} lies in computing the propensities given the parameters.
We have tested \ac{MLE} for estimating $\lambda_j$'s~\cite{DCM2009}.
Though the results with \ac{MLE}~\OurMethod{} are better than the \ac{PBM}-\ac{IPS} in most of the \ac{DCM} simulated click sets, they are worse for $\text{dcm}\_1.0\_0.5$ and not significant for $\text{dcm}\_1.0\_2.0$ (Fig.~\ref{fig:comparison}).
As argued in \cite{chuklin2015click}, \ac{MLE} for \ac{DCM} is based on a simplifying assumption which is not always true.
Our findings coincide with this fact.
Therefore, there is a need for \ac{CLTR}-based algorithms for parameter estimation for \ac{CBM} (see Section \ref{sctn:param_estimate}).
\vspace*{-1mm}
\paragraph{Method Selection}
\label{sctn:goodness}
In order to choose between \ac{PBM}-~and \OurMethod{} for debiasing click logs, a measure that uses historical clicks to validate debiasing models is desired. 
For that, we use click log-likelihood.
Click log-likelihood requires the click probabilities which are computed as the examination probability multiplied by the relevance probability.
The examination probabilities are discussed in Section~\ref{sctn:method}.
For the relevance probabilities we use the output of our ranking function and pass it to different normalizing functions to have a valid probability range.

Our results on the sets presented previously in this section show the followings:
\begin{enumerate*}
	\item $\mathrm{softmax}$ always prefers~\OurMethod~(wrong selection for clicks close to PBM);
	\item $\mathrm{sigmoid}$ always prefers \ac{PBM}-\ac{IPS} (wrong selection for clicks close to CBM); and
	\item exponential $\mathrm{min}\text{-}\mathrm{max}$ selects the better performing approach on the test set (correct selection in both cases).
\end{enumerate*}
We leave more discussions in this regard as future work.

%% file: sections/08-Conclusion.tex

\section{Conclusion}
\ac{PBM} is the default assumption in \ac{IPS}-based \ac{CLTR}.
However, it is unable to properly model the cascade behavior of users.
We raised the question of \ac{PBM} effectiveness in \ac{IPS} unbiased \ac{CLTR} when users click behavior tends to \ac{CBM} \ref{rq:pbm_effectiveness}.
Through a number of experiments, we have answered our \ref{rq:pbm_effectiveness} negatively: \ac{PBM}-\ac{IPS} is not helpful in \ac{CBM} situations and, in our tested cases, there is a gap between its performance and the full-info case.
This answer leads to a more important question:
How to perform \ac{IPS} correction for clicks close to \ac{CBM} \ref{rq:cascade_based_propensity}.
We provided \OurMethod, with closed form formulas for three widely used \acp{CBM}, namely \ac{DCM}, \ac{DBN} and \ac{CCM}.
We have shown the effectiveness of \OurMethod~ on the special case of \ac{DCM}.
Finally, we have given a short discussion about how to select between \ac{PBM}- and \OurMethod~ only by looking at the clicks (and not using the true relevance labels).